\documentclass[aps,pre,reprint,doi=false,isbn=false,url=false,title=true,longbibliography,noeprint,footinbib]{revtex4-1}
\usepackage{graphicx}
\usepackage{amsmath}
\usepackage{array}
\usepackage{amsfonts}
\usepackage{float}
\usepackage{xcolor}
\usepackage[mathscr]{euscript}
\usepackage{catchfile}

\usepackage{array}
\usepackage{makecell}
\def\ba{\begin{eqnarray}}
\def\ea{\end{eqnarray}}
\def\be{\begin{equation}}
\def\ee{\end{equation}}
\def\bm{\begin{math}}
\def\me{\end{math}}

\newcommand{\dummy}

\makeatletter
\newcommand{\fmarki}{*}
\newcommand{\fmarkii}{\ensuremath{\dagger}}
\newcommand{\fmarkiii}{\ensuremath{\ddagger}}
\newcommand{\fmarkiv}{\ensuremath{\mathsection}}
\newcommand{\fmarkv}{\ensuremath{\mathparagraph}}
\newcommand{\fmarkvi}{\ensuremath{\|}}
\newcommand{\fmarkvii}{**}
\newcommand{\fmarkviii}{\ensuremath{\dagger\dagger}}
\newcommand{\fmarkix}{\ensuremath{\ddagger\ddagger}}
                
\def\@fnsymbol#1{{\ifcase#1\or \fmarki\or \fmarkii\or \fmarkiii\or \fmarkiv\or \fmarkv\or \fmarkvi\or \fmarkvii\or \fmarkviii\or \fmarkix \else\@ctrerr\fi}}
\makeatother

\renewcommand{\fmarki}{$\dagger$}
 \renewcommand{\fmarkii}{*}
 \renewcommand{\fmarkiii}{$\ddagger$}
 \renewcommand{\fmarkiv}{a$_4$}
 \renewcommand{\fmarkv}{x$_5$}

\begin{document}

\title{Persistence Properties of a Phase-ordering System with Competing Dynamics}
\author{Shubham Thwal}\email[]{shubhamthwal1@gmail.com}
\author{Suman Majumder}\email[]{smajumder@amity.edu, suman.jdv@gmail.com}
\affiliation{Amity Institute of Applied Sciences, Amity University Uttar Pradesh, Noida 201313, India}
\date{\today}
\begin{abstract}
Understanding the persistence properties of a system with respect to its underlying microscopic dynamics provides important insights into the associated first-passage behavior. Here, we investigate the persistence properties during phase ordering in the two-dimensional ($d=2$) Ising model evolving under competing nonconserved spin-flip and conserved spin-exchange dynamics by means of Monte Carlo simulations at zero temperature. We examine three distinct persistence probabilities: (i) the total persistence probability, defined as the probability that a lattice site has never experienced a change in the sign of the spin residing there; (ii) the spin-flip persistence probability, which exclusively measures the fraction of sites that have never undergone a spin-flip event; and (iii) the composite persistence probability, defined as the fraction of sites that have experienced neither a spin-flip nor a spin-exchange event. In the asymptotic regime, both the total and spin-flip persistence probabilities exhibit identical power-law decay, irrespective of the relative occurrence probability of the spin-flip move $p_r$. The corresponding persistence exponent $\theta_i \approx 0.225$, is found to be consistent with the value reported for systems evolving purely under nonconserved dynamics. We further demonstrate that both persistence measures satisfy the scaling relation $d-d_f^i=\theta_i/\alpha_i$, where $d_f^i$ is the fractal dimension of the corresponding persistence lattice and $\alpha_i\approx 1/2$ characterizes the asymptotic power-law growth of spatially correlated regions of non-persistent spins. In contrast, although the composite persistence probability also exhibits asymptotic power-law decay, both the corresponding persistence exponent $\theta_{\rm c}$ and the fractal dimension $d_f^{\rm c}$ of the persistence lattice show strong dependence on $p_r$. Combined with the presence of a universal growth exponent $\alpha_{\rm c}\approx 1/2$, this leads to the breakdown of the scaling relation among the characteristic exponents for the composite persistence.
\end{abstract}

\maketitle

\section{INTRODUCTION}\label{intro}
The standard protocol to investigate the kinetics of a phase transition is to bring the concerned system out of its current equilibrium state to a new state following an abrupt change in a thermodynamic parameter, e.g., temperature ($T$)  \cite{bray2002,puri_book}. This nonequilibrium process is highlighted via formation and growth of domains, which is referred to as phase ordering or coarsening. It is a scaling phenomenon, characterized by the power-law growth of the average time-dependent length scale or domain size as
\begin{equation}\label{lt_power_law}
 \ell(t) \sim t^{\alpha},
\end{equation}
where $\alpha$ is the growth exponent. In this regard, the nearest neighbor Ising model (NNIM) representing a plethora of such transitions, has been extensively studied in the last 100 years,  both analytically  and by means of Monte Carlo (MC) simulations \cite{Ising_rev1,Ising_rev2}.  While the equilibrium aspects of these transitions are found to have robust universality, the corresponding nonequilibrium dynamics is strongly dependent on the intrinsic transport mechanism considered. For example, during ferromagnetic ordering (achieved using MC simulations of the NNIM using nonconserved dynamics) one observes a domain growth with $\alpha=1/2$ representing the Lifshitz-Cahn-Allen (LCA) law \cite{allen1979}. On the other hand, in a binary mixture phase separation (achieved using MC simulations of the NNIM with conserved dynamics) one gets $\alpha=1/3$ in accordance with the Lifshitz-Slyozov (LS) law \cite{lifshitz1961,lifshitz1962}.
\par
Apart from the scaling behavior associated with domain growth, persistence constitutes another fundamental aspect of the nonequilibrium dynamics of phase transitions, being intimately connected to the first-passage properties of the evolving system. Mathematically, a first-passage time is defined as the earliest time when a stochastic variable changes its state or crosses a specified threshold. Most studies have focused on the persistence probability $P(t)$, defined as the probability that a local degree of freedom has not experienced a first-passage event up to time $t$. For ferromagnetic ordering using the Ising model, this corresponds to the probability that a site has never undergone a spin flip following a quench from a disordered state to an ordered state at $T\equiv 0$. Quantitatively, for a $d$-dimensional hyper cubic system of length $L$, one defines
\begin{equation}\label{persistence_def}
P(t)=1-\frac{N_{\rm f}(t)}{V},
\end{equation}
where $V\equiv L^d$ is the system volume and $N_{\rm f}(t)$ denotes the number of sites which have experienced a spin flip at least once during the time  interval $[0,t]$. Typically, the persistence probability exhibits a power-law decay \cite{Bray2013},
\begin{equation}\label{persitence}
P(t)\sim t^{-\theta},
\end{equation}
where $\theta$ is the persistence exponent. For ferromagnetic ordering (nonconserved dynamics) in $d=1$ theoretical prediction \cite{Derrida1995,Derrida1996} of $\theta=3/8$  is in accordance with the reported simulation result \cite{Derrida1994}. In $d=2$, approximate theory \cite{Majumdar1996} predicts $\theta \approx 0.19$ , whereas the values of $\theta \in [0.19,0.23]$ reported in simulations are slightly higher \cite{Manoj2000,Blanchard2014,Chakraborty2016}. Recent progress in computing facilities has led to the shift of focus on understanding coarsening in more realistic and complex systems. In this regard, there are studies that has also explored the persistence properties of the long-range Ising model \cite{Christiansen2021}.

\par
For the NNIM with purely conserved dynamics, the conventional definition of persistence fails to capture the intrinsic spin-exchange mechanism and consequently exhibits rather a featureless behavior. In contrast, for systems where conserved and nonconserved dynamics coexist \cite{glotzer1994,puri1994,glotzer1995,zhu2002}, the standard persistence definition still remains meaningful. The interest in such systems with mixed dynamics stems from the nonequilibrium behavior arising from the competition between phase segregation and interconversion reaction in binary mixtures. Examples include isomeric mixtures, where the components simultaneously inter convert and segregate, often leading to enrichment of one species \cite{soai1995,tran1996reaction,Qui1997,lombardo2009,shumovskyi2021}. The interplay of these processes also provides a useful framework for understanding collective phenomena such as cell-adhesion dynamics \cite{blom2021}, opinion-driven social segregation in the Schelling model \cite{barret2024}, and chemically active droplets relevant to intra-cellular organization \cite{weber2019}. 
\par
The phase-ordering dynamics of the NNIM with mixed dynamics at finite $T$ has previously been investigated using Monte Carlo simulations \cite{grandi1997,huang1998scaling,szolnoki2000,thwal1}. More recently, we examined how the interplay between nonconserved spin-flip dynamics and conserved spin-exchange dynamics influences the scaling behavior of the growing length scale \cite{thwal2}. Although both mechanisms operate simultaneously, the system ultimately evolves toward a state dominated by a single spin species, rendering the late-time dynamics effectively nonconserved. Nevertheless, the competition between the two dynamical processes at intermediate times gives rise to considerably richer behavior than either pure dynamics alone, leading to a crossover phenomenon in the growth kinetics. Thus, at $T=0$, such a system is expected to exhibit intriguing dynamical behavior, particularly in how the competing conserved and nonconserved processes influence the persistence properties of an effectively nonconserved system. Despite this, persistence in the NNIM with competing dynamics has, to the best of our knowledge, remained unexplored so far.

\par
Inspired by the above discussion on mixed dynamics and their rich nonequilibrium behavior, in this work we investigate the persistence properties of the NNIM with competing conserved and nonconserved dynamics at $T=0$. In particular, we study not only the conventional total spin-flip persistence, but also the persistence associated exclusively with spin-flip events and a composite persistence that incorporates the combined effects of both dynamical processes. We further analyze the spatial organization of the corresponding persistence lattices by examining their growth dynamics, scaling behavior, and fractal characteristics. Scaling relations connecting the persistence-decay exponent, growth exponents, and fractal dimensions are also carefully tested to understand how the interplay between the competing dynamics modifies the nonequilibrium evolution of the system.
\par
The rest of the paper is organized in the following way. The next section discusses the model and method of simulations. Section\ \ref{results} presents the results. Finally, Sec.\ \ref{conclusion} provides the conclusive remarks and a future outlook.

\section{Model and Method}\label{methods}
The Hamiltonian for the NNIM is given as 
\begin{eqnarray}\label{Ising}
 H = -J\sum_{\langle ij \rangle}S_iS_j,
\end{eqnarray}
where the spin $S_i$ can take up values $=+1~\rm{or}~-1$,  and $J~(=1)$ represents the strength of interaction. The sign $\langle \dots \rangle$ denotes nearest-neighbor interactions among the spins. We use a square lattice with periodic boundary condition (pbc) in all directions. The critical temperature of the NNIM in a square lattice with pbc is $T_c=2J/k_B\ln(1+\sqrt{2})$,
where $k_B$ is the Boltzmann constant. 
\par
 We perform MC simulations of the above model with moves that incorporates both the nonconserved and conserved dynamics. For the nonconserved dynamics we employ the Glauber spin-flip move  \cite{glauber1963,barkema_book,landau_book}, where one randomly picks a lattice site and attempts to flip the sign of the spin residing there. The conserved dynamics  is introduced via the Kawasaki spin-exchange move \cite{kawasaki1966,barkema_book,landau_book}. There one randomly picks up a pair of nearest-neighbor sites and exchange the spins at those positions. As we are interested only in the $T=0$ dynamics, we accept an attempted MC move if the energy change $\Delta E \le 0$, which corresponds to the Metropolis criterion \cite{barkema_book,landau_book}.
\par
As the initial condition, we prepare a homogeneous $50{:}50$ mixture of $+$ and $-$ spins by randomly assigning them on the lattice sites, corresponding to a high-temperature disordered state. The system is then quenched to $T=0$, where the relative competition between nonconserved Glauber spin-flip dynamics and conserved Kawasaki spin-exchange dynamics is controlled through the probability $p_r$ of the nonconserved move. This means that at each attempted MC move, a spin-flip move is chosen with probability $p_r$, while a spin-exchange move is attempted with a  probability $1-p_r$. Simulations are performed for $p_r \in [0.001,1]$  with  $L \in [64,256]$. Time is measured in units of one Monte Carlo sweep (MCS) that consists of $L^2$ attempted MC moves. The simulations are run up to a maximum time ranging between $10^5$ to $10^7$ MCS depending on $L$ and $p_r$. All results, unless otherwise stated, are averaged over $20$ independent initial configurations generated using different random number seeds. The important observables necessary for our analyses are defined subsequently wherever they appear for the first time.
\begin{figure}[t!]
\centering
\includegraphics*[width=0.5\textwidth]{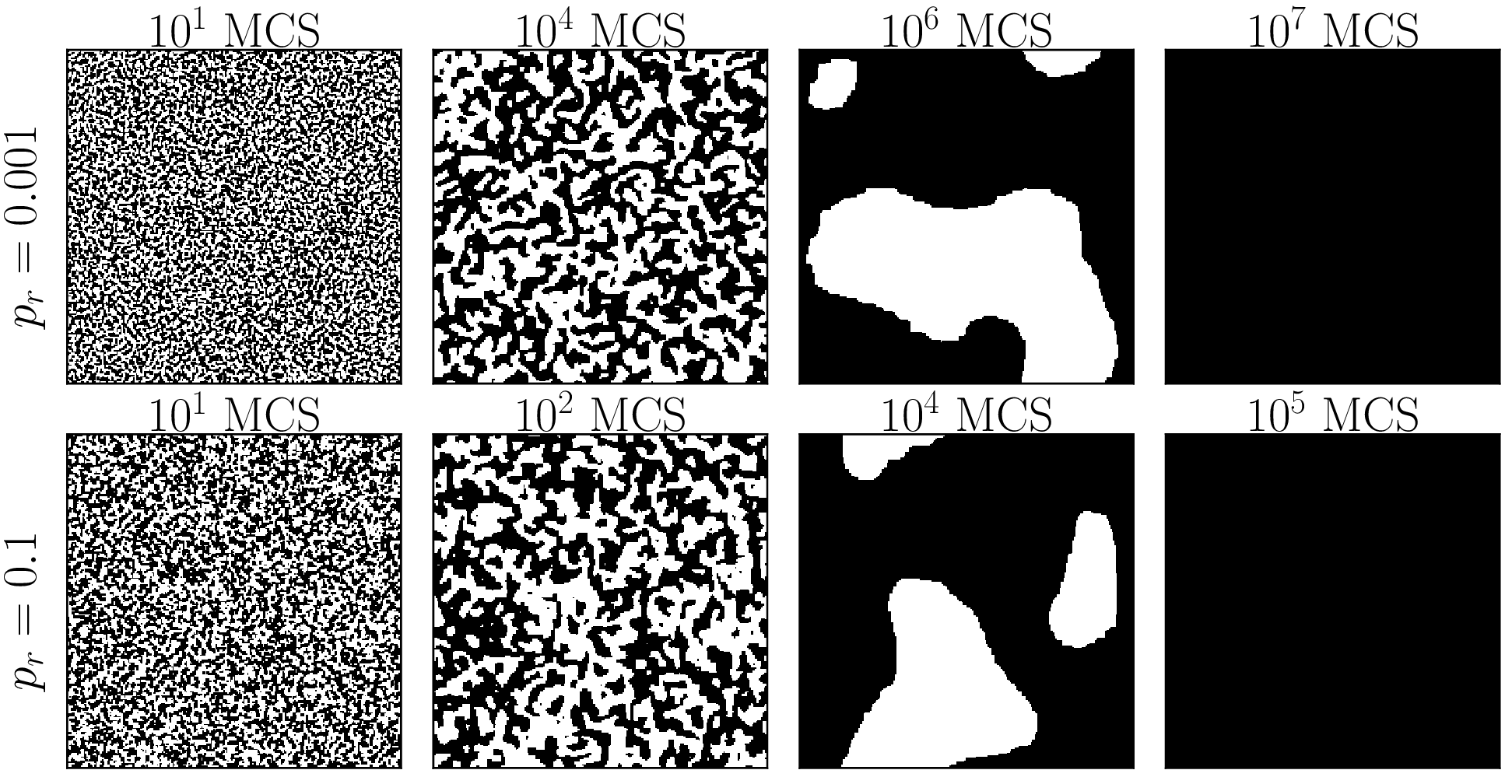}\\
\caption{\label{snap_original}Snapshots  of the original lattice  at different times depicting the evolution of a system having $L=256$ during phase ordering with mixed conserved and nonconserved dynamics at $T=0$, for two values of $p_r$. The up spins are marked black and the down spins are left unmarked.}
\end{figure}
\section{Results}\label{results}

\subsection{Dynamics of the Original Lattice}\label{original_lattice}
Before exploring the persistence properties, we first examine the scaling behavior and growth dynamics of the original lattice. Accordingly, in Fig.\ \ref{snap_original}, we present representative snapshots illustrating the time evolution during phase ordering at $T=0$ for two different values of $p_r$. In both cases, domains of like spins are seen to form and grow with time. For all values of $p_r$, most simulation runs eventually evolve into fully ordered states consisting entirely of either up or down spins. In a few cases, however, the system becomes trapped in metastable slab-like configurations. The snapshots also reveal clear differences in the ordering time scales for the cases, indicating that the process becomes faster with increasing  $p_r$. 
\begin{figure}[t!]
\centering
\includegraphics*[width=0.5\textwidth]{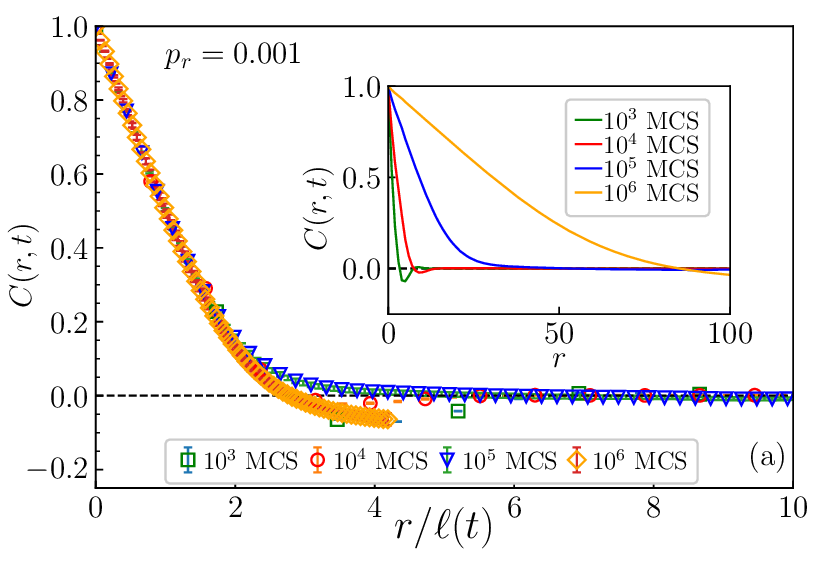}\\
\vskip 0.2cm
\includegraphics*[width=0.5\textwidth]{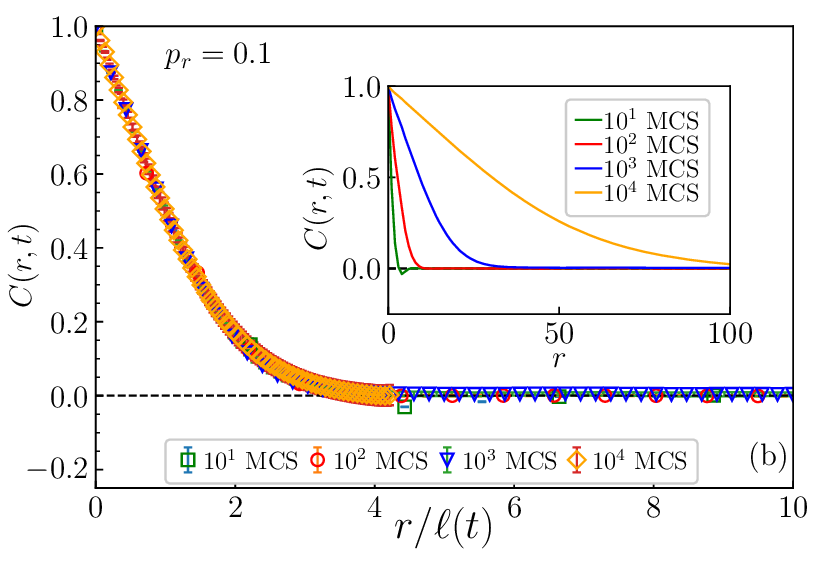}\\
\vskip 0.2cm
\includegraphics*[width=0.5\textwidth]{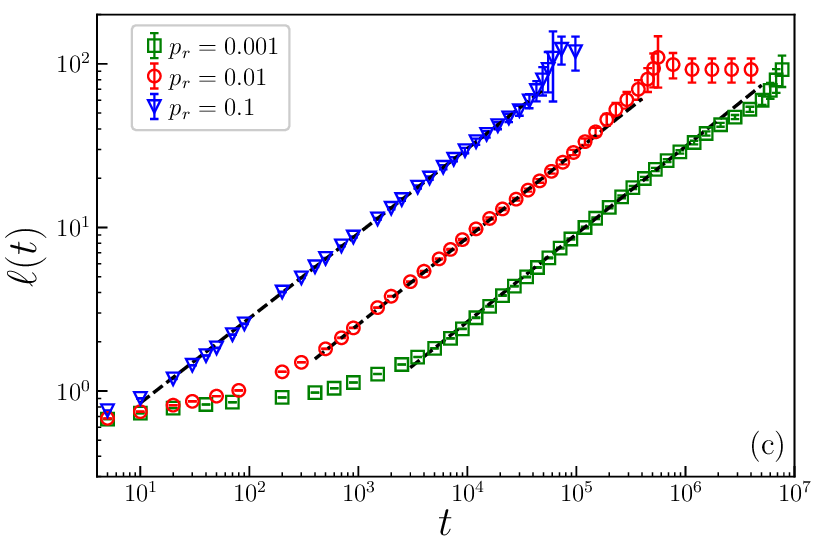}\\
\caption{\label{scaling_original} Demonstration of the scaling behavior encrypted in Eq.\ \eqref{scaling_Cr} for (a) $p_r=0.001$  and (b) $0.1$. The main frame shows the plot of $C(r,t)$ as a function $r/\ell(t)$ at different times, as indicated. The inset shows the corresponding unscaled plots. (c) Double-log plots of the time dependence of the average domain length $\ell(t)$ for three different values of $p_r$ for a system of length $L=256$. The black dashed lines represent the respective power-law fits using the ansatz in Eq.\ \eqref{lt_fit}.}
\end{figure}
\par
To verify domain growth as a scaling phenomenon, we next calculate the two-point equal-time order parameter correlation function
\begin{equation}\label{Cr_func}
 C(r,t)=\langle S_i S_j\rangle - \langle S_i\rangle \langle S_j\rangle,
\end{equation}
where $r=|i-j|$ denotes the distance between the lattice sites $i$ and $j$. As $t$ increases, $C(r,t)$ decays to zero at progressively larger values of $r$, indicating the presence of a growing length scale that characterizes the average domain size $\ell(t)$. We estimate $\ell(t)$ using the criterion
\begin{equation}\label{lt_Cr}
 C[r=\ell(t),t]=\frac{1}{2}C(0,t).
\end{equation}
For a scaling phenomenon, $C(r,t)$ together with $\ell(t)$ is expected to obey the relation \cite{bray2002,puri_book}
\begin{equation}\label{scaling_Cr}
C(r,t)\equiv \tilde{C} \left(r/\ell(t)\right),
\end{equation}
where $\tilde{C}(x)$ is a time-independent scaling function. The above scaling behavior is demonstrated in Figs.\ \ref{scaling_original}(a) and (b) for the systems shown in Fig.\ \ref{snap_original}. The main panels display $C(r,t)$ as a function of the scaled variable $r/\ell(t)$, while the corresponding unscaled plots are shown in the insets. The collapse of data obtained at different times onto a single master curve clearly confirms the scaling form stated in Eq.\ \eqref{scaling_Cr}. For $p_r=0.001$, the early-time data exhibit a substantial decay below zero at large values of $r$. This behavior is characteristic of systems evolving predominantly through conserved dynamics, which is indeed the case at early times for small $p_r$. In contrast, for larger values of $p_r$, both conserved and nonconserved dynamics contribute significantly from the very beginning, rendering the overall dynamics effectively nonconserved. Consequently, the negative decay of $C(r,t)$ at large $r$ is barely observed in this case.
\par
In Fig.\ \ref{scaling_original}(c), we demonstrate the power-law scaling behavior of the time dependence of the average domain size $\ell(t)$. The data exhibit an initial transient regime before eventually crossing over to a late-time power-law growth. The duration of this transient regime decreases with increasing $p_r$. Similar behavior has also been reported for the dynamics at finite temperatures \cite{thwal2}. To quantify the late-time growth behavior, we fit the data using the ansatz
\begin{equation}\label{lt_fit}
\ell(t)=At^{\alpha}. 
\end{equation}
The fitting is performed by excluding both the early-time transient regime and the late-time finite-size regime where $\ell(t)$ saturates. The resulting estimates of the growth exponent $\alpha$ are $0.536(2)$, $0.530(3)$, and $0.518(3)$ for $p_r=0.001$, $0.01$, and $0.1$, respectively. These values are in reasonable agreement with the LCA prediction for a purely nonconserved dynamics. The dashed lines passing through the data points represent the corresponding fitted curves. 
\begin{figure}[t!]
\centering
\includegraphics*[width=0.5\textwidth]{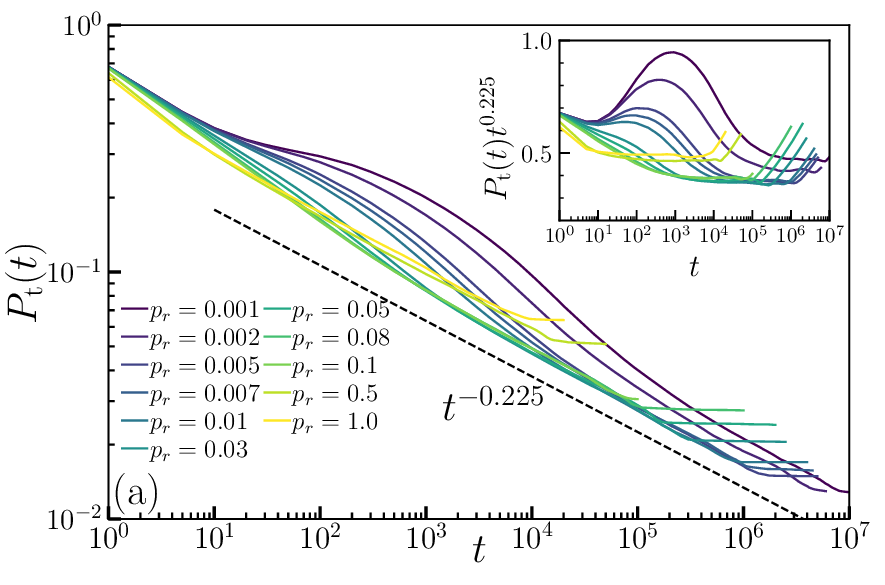}\\
\vskip 0.2cm
\includegraphics*[width=0.5\textwidth]{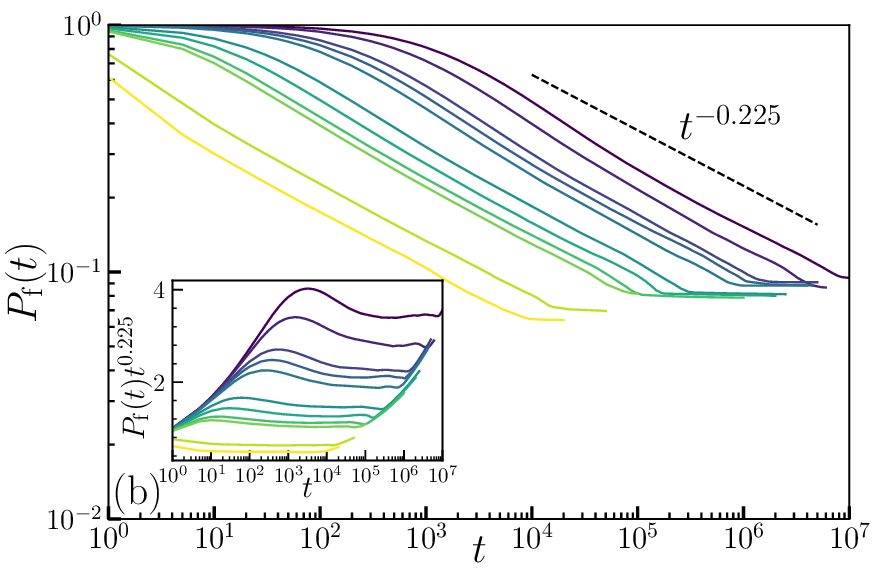}\\
\caption{\label{persistence_prob}(a) The main panel shows log-log plots of the total persistence probability $P_{\rm t}(t)$ as a function of time for different values of $p_r$. The dashed line demonstrates the consistency of the late-time data with a power-law decay characterized by the exponent $\theta_{\rm t}=0.225$. The inset displays the time dependence of the quantity $P_{\rm t}(t)t^{0.225}$ to further illustrate the agreement of the late-time behavior with the expected power-law scaling. (b) The same set of plots for the spin-flip persistence probability $P_{\rm f}(t)$. In all cases, the data correspond to a system size $L=256$.}
\end{figure}

\subsection{Total and Spin-Flip Persistence Properties}\label{persistence_lattice}
Next, we turn our attention to the persistence properties of the phase-ordering system evolving under mixed dynamics. As already discussed in Sec.\ \ref{intro}, persistence is traditionally defined for nonconserved dynamics as the probability that a lattice site has never experienced a change in sign of the residing spin up to a given observation time. However, for a system evolving under mixed dynamics, this conventional definition does not distinguish whether a sign change occurs due to a nonconserved spin-flip move or a conserved spin-exchange move. Therefore, we refer to this quantity as the total persistence, denoted by $P_{\rm t}(t)$. We also calculate the probability that a site has never undergone a spin-flip event up to the observation time, irrespective of any spin-exchange processes. We refer to this quantity as the spin-flip persistence, denoted by $P_{\rm f}(t)$. In the main panels of Figs.\ \ref{persistence_prob}(a) and (b), we present the time dependence of the corresponding persistence probabilities. In the long-time limit, both $P_{\rm t}(t)$ and $P_{\rm f}(t)$ exhibit linear behavior on a double-logarithmic scale, indicating a power-law decay of the form
\begin{equation}\label{P_power_law}
 P_{i}(t) \sim t^{-\theta_i}; ~i\equiv {\rm t,~f}.
\end{equation}
The consistency of the late-time data for both $P_{\rm t}(t)$ and $P_{\rm f}(t)$ with the corresponding dashed lines in the main panels suggests that $\theta_{\rm t}=\theta_{\rm f}\approx 0.225$. This value is in good agreement with the persistence exponent reported for systems evolving purely under nonconserved dynamics. For larger values of $p_r$, both $P_{\rm t}(t)$ and $P_{\rm f}(t)$ directly approach the late-time scaling regime. In contrast, for smaller $p_r$, $P_{\rm t}(t)$ exhibits an initial transient regime before entering the asymptotic power-law decay. A more pronounced transient behavior is observed for $P_{\rm f}(t)$, where the data remain nearly constant at early times and begin to decay only after a certain crossover time. This behavior can be attributed to the fact that for smaller values of $p_r$, spin-flip events are relatively rare. Consequently, during the initial period, $P_{\rm t}(t)$ decreases due to spin-exchange events, whereas $P_{\rm f}(t)$ remains nearly unchanged. Notably, the transient period observed in $P_{\rm f}(t)$ approximately coincides with the initial transient regime of $P_{\rm t}(t)$. With increasing $p_r$, the duration of this transient regime decreases and eventually becomes negligible for sufficiently large values of $p_r$. 
\begin{figure}[t!]
\centering
\includegraphics*[width=0.5\textwidth]{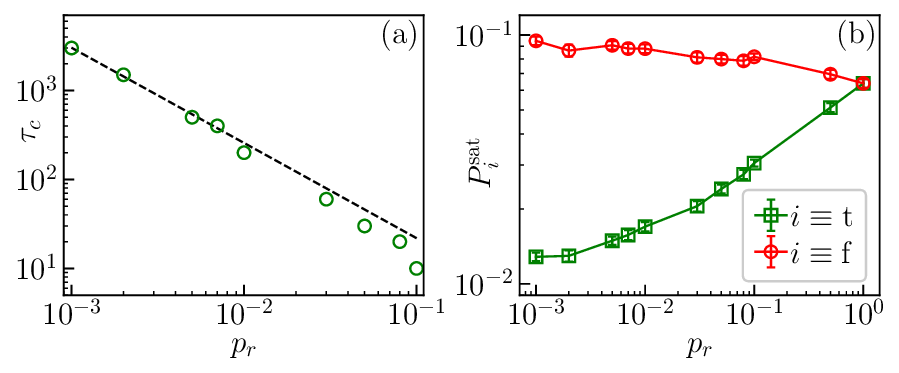}\\
\caption{\label{psat_vs_pr} (a) Scaling of the crossover time $\tau_c$ as a function of $p_r$ for phase ordering system of size $L=256$. The dashed line is a fit to form  in Eq.\ \ref{tc_vs_pr}. (b) Variation of the saturated value of persistence probabilities as a function of $p_r$.}
\end{figure}
\par
To further demonstrate the consistency of the late-time behavior with the expected power-law scaling, in the insets of Figs.\ \ref{persistence_prob}(a) and (b) we plot the time dependence of the quantity $P_{i}(t)t^{0.225}$. In the long-time limit, the data clearly exhibit a plateau, confirming the expected asymptotic scaling behavior. Importantly, the data for $P_{\rm f}(t)t^{0.225}$ display the presence of a peak or hump, the prominence of which decreases with increasing $p_r$. The position of this peak approximately coincides with the time at which $P_{\rm f}(t)$ enters the asymptotic power-law scaling regime. We therefore use the location of this peak to estimate the crossover time $\tau_c$, which characterizes the onset of spin-flip-dominated dynamics over spin-exchange processes during phase ordering.
In principle, $\tau_c$ may also be extracted from the corresponding features in $P_{\rm t}(t)$. However, for larger values of $p_r$, the estimate obtained from $P_{\rm f}(t)$ is found to be more reliable. In Fig.\ \ref{psat_vs_pr}(a), we show the variation of $\tau_c$ as a function of $p_r$. The approximately linear behavior observed on a double-logarithmic scale suggests an underlying power-law decay, which we quantify using the ansatz
\begin{equation}\label{tc_vs_pr}
 \tau_c=Kp_r^{-x},
\end{equation}
where $K$ is a nontrivial prefactor and $x$ is the corresponding decay exponent. A fitting exercise using the above ansatz yields $K=1.87(31)$ and $x=1.07(3)$.
We also observe from the main panels of Fig.\ \ref{persistence_prob} that the saturation values of $P_{\rm t}(t)$ and $P_{\rm f}(t)$ exhibit distinct dependence on $p_r$. As shown in Fig.\ \ref{psat_vs_pr}(b), the saturation value of the total persistence $P_{\rm t}^{\rm sat}$ increases steadily with increasing $p_r$. In contrast, the corresponding saturation value for the spin-flip persistence $P_{\rm f}^{\rm sat}$ decreases gradually with increasing $p_r$.
\begin{figure}[t!]
\centering
\includegraphics*[width=0.5\textwidth]{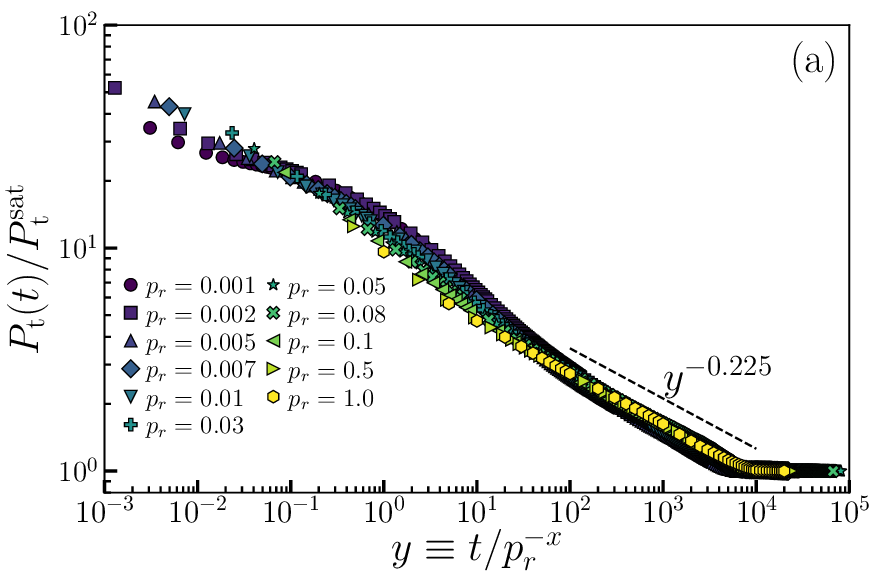}\\
\vskip 0.1cm
\includegraphics*[width=0.5\textwidth]{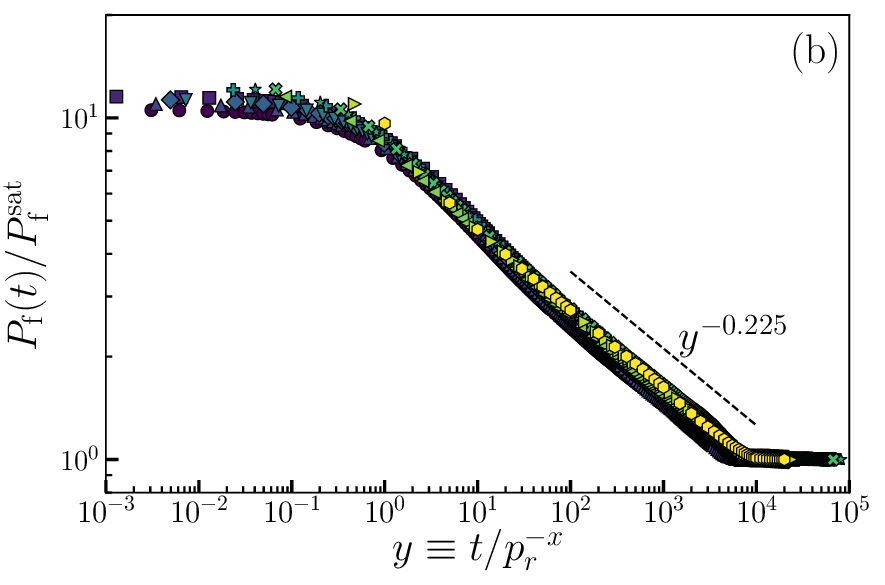}\\
\caption{\label{scaling_persistence} ((a) Double-logarithmic plots of the scaled total persistence probability $P_{\rm t}(t)/P_{\rm t}^{\rm sat}$, as a function of the scaling variable $y\equiv t/\tau_c$ for different values of $p_r$, obtained for a system of size $L=256$. (b) Corresponding plots for the scaled spin-flip persistence probability $P_{\rm f}(t)/P_{\rm f}^{\rm sat}$. The dashed lines in both panels represent the expected power-law decay $\sim y^{-0.225}$.}
\end{figure}

\par
To examine the possibility of a universal dynamic scaling of $P_i$ for different values of $p_r$, we propose the scaling form
\begin{equation}\label{scaling_P}
 P_{i}(t)=P_{i}^{\rm sat}f(y);~{\rm where}~y=\frac{t}{\tau_c}\equiv \frac{t}{p_r^{-x}}.
\end{equation}
In Eq.\ \eqref{scaling_P}, $f(y)$ is a scaling function independent of $p_r$, and $y$ is the corresponding scaling variable. If such dynamic scaling exists, one should observe a collapse of data obtained for different values of $p_r$ upon plotting $P_{i}(t)/P_{i}^{\rm sat}$ as a function of $y$. In the long-time limit $y\gg 1$, the scaling function is expected to follow
\begin{equation}\label{scaling_func}
 f(y)\sim y^{-\theta_i}.
 \end{equation}
For the total persistence, the regime $y\ll 1$ corresponds to dynamics dominated by spin-exchange processes, and therefore one expects a relatively featureless behavior in this limit. In contrast, for the spin-flip persistence, while the large-$y$ regime is expected to follow the asymptotic behavior given in Eq.\ \eqref{scaling_func}, in the limit $y\ll 1$ the scaling function is expected to remain approximately constant. Figure\ \ref{scaling_persistence} presents the corresponding scaling analyses for both the total persistence and spin-flip persistence. The data for $P_{\rm t}(t)$ shown in Fig.\ \ref{scaling_persistence}(a), although not displaying an excellent collapse over the entire range of $y$, exhibit a reasonably good collapse in the asymptotic regime $y\gg 1$, where the behavior is found to be consistent with Eq.\ \eqref{scaling_func}. On the other hand, the data for $P_{\rm f}(t)$ show a reasonably good collapse over the entire range of $y$, while also displaying the expected plateau behavior at small $y$ and the power-law decay at large $y$ as predicted in Eq.\ \eqref{scaling_func}.
\begin{figure}[t!]
\centering
\includegraphics*[width=0.5\textwidth]{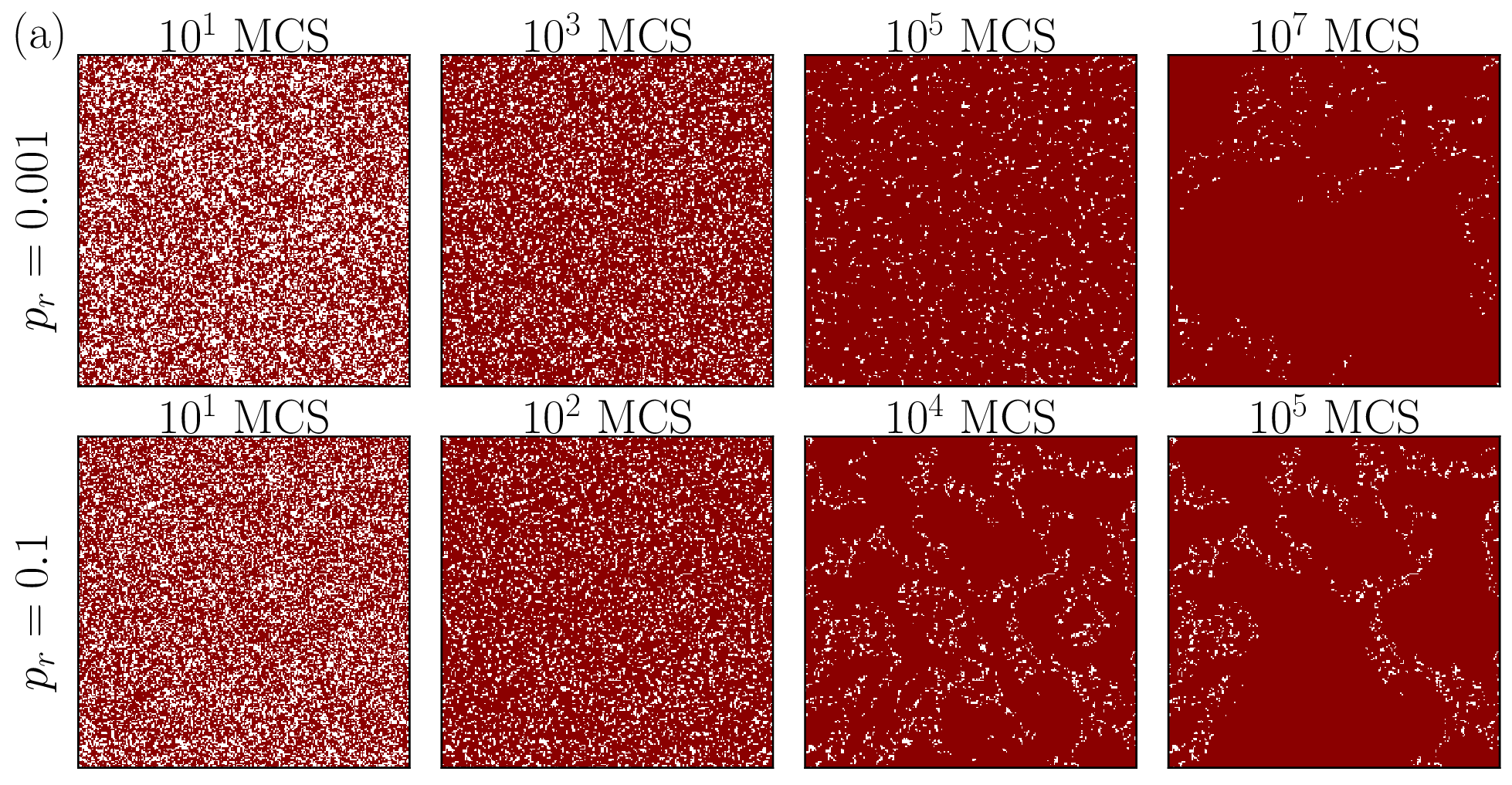}\\
\vskip 0.1cm
\includegraphics*[width=0.5\textwidth]{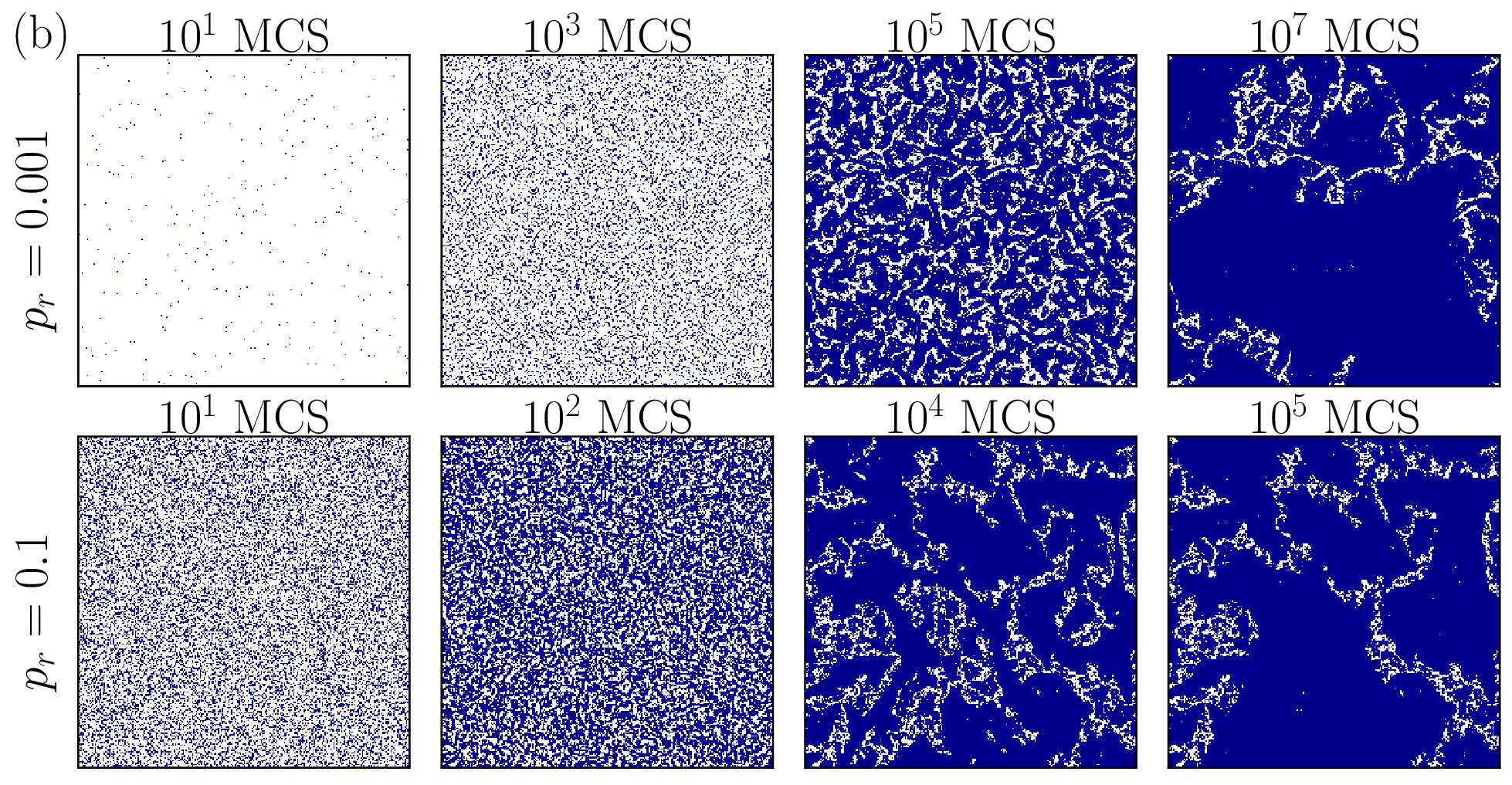}\\
\caption{\label{snap_persistence}  (a) Time evolution of the total persistence lattice for the same system presented in Fig.\ \ref{snap_original}. For estimating the total persistence, a site is said to be persistent if the sign of the spin at that site has never changed until the observation time $t$. Here the sites that have undergone sign change at least once are marked red and the persistent sites are left unmarked. (b) Corresponding time evolution of the spin-flip persistence lattice. In this case, a site is considered to be persistent if it has never encountered a successful spin-flip move until the observation time. In the snapshots the sites that have undergone at least one spin flip are marked blue and the persistence sites are left unmarked.}
\end{figure}
\par
Another important aspect of persistence concerns the morphological properties of the corresponding persistence lattice. To investigate this, we define an order parameter $\rho_i(\mathbf{x}_i,t)$, where $i$ denotes either the total or spin-flip persistence and $\mathbf{x}_i$ represents the lattice position vector. At a given time, if a site remains persistent, then $\rho_i(\mathbf{x}_i,t)=1$; otherwise, $\rho_i(\mathbf{x}_i,t)=0$. Representative snapshots at different times for two different values of $p_r$ are shown in Figs.\ \ref{snap_persistence}(a) and (b) for the total and spin-flip persistence lattices, respectively. In these snapshots, only sites with $\rho_i(\mathbf{x}_i,t)=0$ are marked using colors. In both cases, the evolving morphologies indicate the growth of spatially correlated non-persistent regions, which appear to acquire a fractal-like structure at late times. The qualitative evolution for different values of $p_r$ follows a trend similar to that observed for the original spin lattice shown in Fig.\ \ref{snap_original}.
\par
The spatial correlations that develop over time can be characterized using the correlation function \cite{Manoj2000,manoj2000scaling}

\begin{equation}\label{corr_pers_def}
 D_i(r,t)=\frac{\langle \rho_{i}(\mathbf{x}_i,t)\rho_{i}(\mathbf{x}_i+\mathbf{r},t)\rangle}{\langle \rho_{i}(\mathbf{x}_i,t) \rangle};~ i\equiv {\rm t,~f},
\end{equation}
where $\mathbf{r}$ denotes the separation vector between two lattice sites. Here, the angular brackets represent averages over independent simulation runs. By definition, $\langle \rho_i(\mathbf{x}_i,t)\rangle=P_i(t)$. For a system evolving under purely nonconserved dynamics, $D_i(r,t)$ is known to obey the following dynamic scaling form \cite{Manoj2000,manoj2000scaling}
\begin{equation}\label{corr_scaling_func}
 D_i(r,t)=P_i(t)\mathcal{F}_i(y);~i\equiv {\rm t,~f},
\end{equation}
where $\mathcal{F}_i(y)$ is the corresponding scaling function. In Eq.\ \eqref{corr_scaling_func}, the scaling variable is defined as

\begin{equation}\label{scaling_y}
 y=\frac{r}{\ell_p^i(t)}; ~i\equiv {\rm t,~f},
\end{equation}
where $\ell_p^i(t)$ denotes the characteristic length scale at time $t$, associated with the growth of correlated non-persistent regions, as also evident from Fig.\ \ref{snap_persistence}. Numerically we estimate $\ell_p^i(t)$ from the value of $r$ where $D_i(r,t)\approx P_i(t)$. Typically, $\ell_p^i(t)$ is expected to exhibit a power-law growth of the form

\begin{equation}\label{lp_law}
\ell_p^i(t)=A_i t^{\alpha_i}.
\end{equation}
For purely nonconserved dynamics, the growth exponent $\alpha_i$ has been found to be consistent with the domain growth exponent $\alpha$ of the original spin lattice \cite{Manoj2000,Chakraborty2016}.
\begin{figure}[b!]
\centering
\includegraphics*[width=0.5\textwidth]{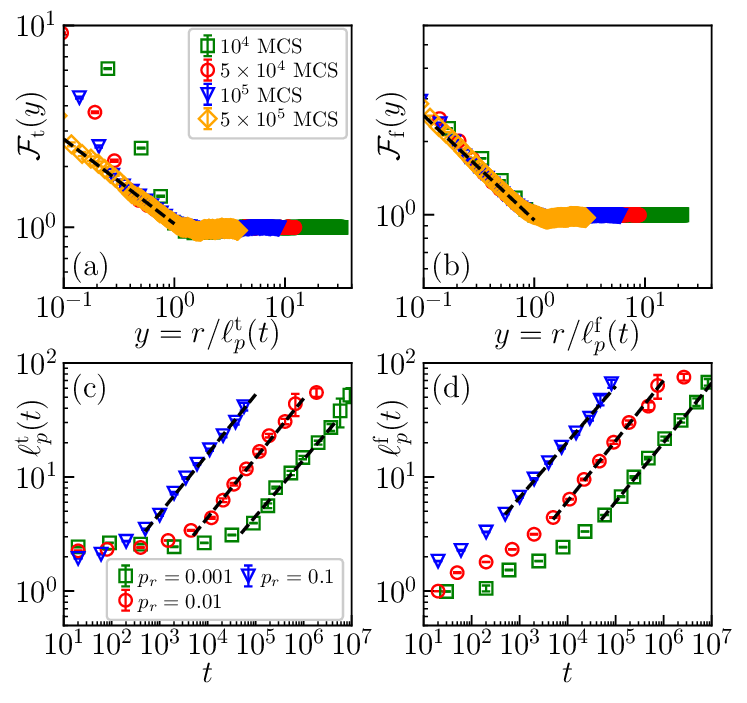}\\
\caption{\label{length_persistence} (a) Double-logarithmic plots of the scaling function $\mathcal{F}_{\rm t}(y)$ for the total persistence at different times for a system evolving with $p_r=0.01$. The dashed line represents a fit to the power-law decay with exponent $\kappa=0.42(2)$, characterizing the small-$y$ behavior described in Eq.\ \eqref{fy_behavior}. (b) Corresponding plots of the scaling function for the spin-flip persistence. Here, the dashed line represents a power-law decay with $\kappa=0.43(2)$. (c) Double-logarithmic plots of the time dependence of the average length scale $\ell_p^{\rm t}(t)$ characterizing the growth of structures in the total persistence lattice, for three different values of $p_r$. The dashed lines represent the corresponding fits using Eq.\ \eqref{lp_law}, yielding $\alpha_{\rm t}=0.496(7)$, $0.522(5)$, and $0.523(4)$ for $p_r=0.001$, $0.01$, and $0.1$, respectively. (d) Corresponding plots of the characteristic length scale for the spin-flip persistence lattice. The dashed lines again represent power-law fits with exponents $\alpha_{\rm f}=0.521(4)$, $0.529(2)$, and $0.487(2)$ for $p_r=0.001$, $0.01$, and $0.1$, respectively. In all cases, the system size is fixed at $L=256$.}
\end{figure}
\par
The scaling function in Eq.\ \eqref{corr_scaling_func} is expected to exhibit the following asymptotic behavior \cite{Manoj2000,manoj2000scaling}
\begin{equation}\label{fy_behavior}
\mathcal{F}_i(y)=\frac{D_i(r,t)}{P_i(t)}=
\begin{cases}
y^{-\kappa}, & y\ll 1,\\
1, & y\gg 1.
\end{cases}
\end{equation}
The decay exponent $\kappa=d-d_f^i$ is related to the fractal dimension $d_f^i$ of the persistence lattice. Now, considering the small-$y$ behavior of $\mathcal{F}_i(y)$, and combining Eqs.\ \eqref{P_power_law} and \eqref{lp_law}, one obtains $
\kappa\alpha_i=\theta_i,$ or equivalently,

\begin{equation}\label{scaling_relation}
 d-d_f^i=\frac{\theta_i}{\alpha_i}.
\end{equation}
The above relation has been shown to hold for both the NNIM \cite{Chakraborty2016} and the long-range Ising model \cite{Christiansen2021} evolving under purely nonconserved dynamics.

\par
In Figs.\ \ref{length_persistence}(a) and (b), we present representative plots of $\mathcal{F}_i(y)$ for $p_r=0.01$ to examine the scaling behavior embodied in Eq.\ \eqref{fy_behavior} for the total and spin-flip persistence, respectively. The data for the total persistence shown in Fig.\ \ref{length_persistence}(a) reveal that the early-time curves do not overlap with those at later times. This can be attributed to the very fact that, at early times, the dynamics is dominated by conserved spin-exchange moves, whereas at later times the nonconserved single spin-flip moves become dominant. Nevertheless, fitting the latest-time data to the power-law form given in Eq.\ \eqref{fy_behavior} yields $\kappa=0.42(2)$, corresponding to a fractal dimension $d_f^{\rm t}=1.58(2)$. This value is consistent with that reported for systems evolving purely under nonconserved dynamics \cite{Chakraborty2016}. The corresponding results for the spin-flip persistence are shown in Fig.\ \ref{length_persistence}(b), where the data exhibit almost a perfect overlap across different times. Performing a similar fitting analysis using the latest-time data yields $\kappa=0.43(2)$, which implies $d_f^{\rm f}=1.57(2)$, perfectly agreeing with that obtained for the total persistence lattice. We have also analyzed the scaling function $\mathcal{F}_i(y)$ for other values of $p_r$, which yield compatible estimates of the fractal dimension within the range $d_f^i \in [1.51,1.60]$.

\par
In order to examine the validity of the scaling relation in Eq.\ \eqref{scaling_relation}, we next present the time dependence of the characteristic length scale $\ell_p^i(t)$ in Figs.\ \ref{length_persistence}(c) and (d), and quantify the corresponding growth exponent $\alpha_i$ for three different values of $p_r$. The data for both the total and spin-flip persistence lattices exhibit similar behavior, characterized by an initial transient slow-growth regime followed by an asymptotic power-law growth at later times. As in the case of the original spin lattice, the transient regime here is also associated with growth dominated by conserved dynamics. Consequently, its duration increases as $p_r$ decreases. Fitting the data in Fig.\ \ref{length_persistence}(c) to the ansatz in Eq.\ \eqref{lp_law} yields $\alpha_{\rm t}=0.496(7)$, $0.522(5)$, and $0.523(4)$ for $p_r=0.001$, $0.01$, and $0.1$, respectively. Similarly, fitting the data in Fig.\ \ref{length_persistence}(d) provides $\alpha_{\rm f}=0.521(4)$, $0.529(2)$, and $0.487(2)$ for $p_r=0.001$, $0.01$, and $0.1$, respectively. Therefore, the growth exponents for both persistence lattices are consistent, within numerical uncertainty, with the LCA prediction. On recalling that we have obtained $\theta_i=0.225$, and then combining it with the estimated values of $d_f$ and $\alpha_i$,  it is evident that the scaling relation given in Eq.\ \eqref{scaling_relation} remains well satisfied even in the present system evolving under mixed conserved and nonconserved dynamics. 
\subsection{Composite Persistence Properties}\label{composite}
As already mentioned, the total persistence is a direct extension of the conventional definition of persistence used for spin systems evolving under purely nonconserved dynamics. On the other hand, we introduced the spin-flip persistence to isolate the sole effect of nonconserved moves in a system evolving under mixed dynamics. However, neither of these quantities explicitly identifies the sites that remain completely unaffected by both microscopic dynamical processes. 
To probe this aspect, we introduce a composite persistence probability, $P_{\rm c}(t)$, defined as the fraction of lattice sites that have neither experienced a spin-flip event nor undergone a spin-exchange move with neighboring sites up to time $t$. By construction, $P_{\rm c}(t)$ provides a direct quantitative measure of completely inactive sites that remain untouched by any of the underlying microscopic dynamics.
\begin{figure}[t!]
\centering
\includegraphics*[width=0.5\textwidth]{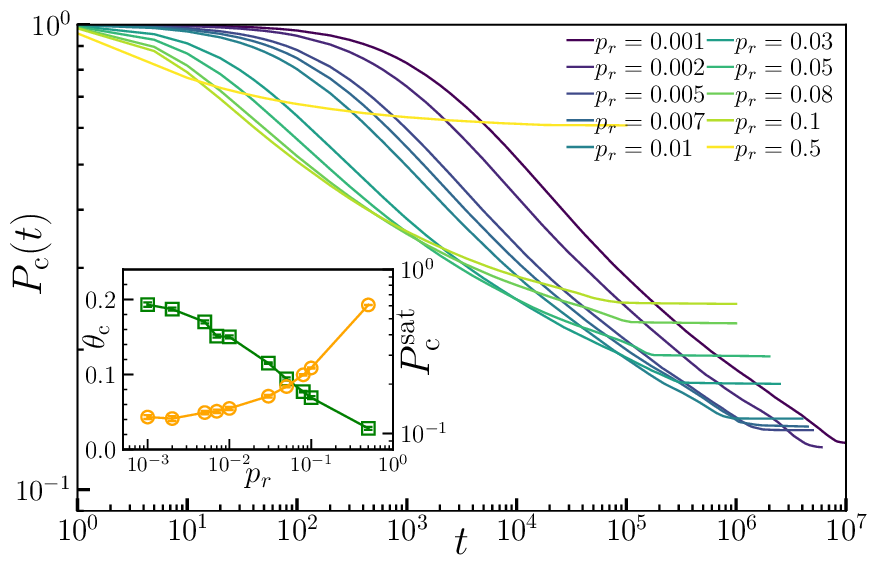}\\
\caption{\label{persistence_composite} The main panel shows the time dependence of the composite persistence probability $P_{\rm c}(t)$ for different values of $p_r$ in a system of size $L=256$ evolving under mixed dynamics. The inset presents the variation of the estimated asymptotic power-law decay exponent (squares), along with the corresponding saturation value $P_{\rm c}^{\rm sat}$ (circles) of $P_{\rm c}(t)$, as a function of $p_r$.}
\end{figure}
\begin{table}[b!]
\centering
\caption{Details of the fitting exercise performed on the long-time data of the composite persistence probability $P_{\rm c}(t)$ using Eq.\ \eqref{P_power_law}.}
\label{tab1}
\begin{tabular}{lll} 
\hline
\hline
~~~$p_r$~~~&~~~~~fitting $t$ range~~~~~~~~~~&~~~~~$\theta_{\rm c}$ \\
\hline
$0.001$ & $[1\times 10^5:3\times 10^6]$ & $0.193(3)$ \\
$0.002$ & $[1\times 10^5:2\times 10^6]$ & $0.187(3)$ \\
$0.005$ & $[8\times10^4:8\times 10^5]$ & $0.170(2)$ \\
$0.007$ & $[7\times10^4:7\times 10^5]$ & $0.151(2)$ \\
~~$0.01$ & $[5\times10^4:5\times 10^5]$ & $0.150(2)$ \\
~~$0.03$ & $[2\times10^4:2\times 10^5]$ & $0.115(2)$ \\
~~$0.05$ & $[8\times10^3:9\times 10^4]$ & $0.094(2)$ \\
~~$0.08$ & $[7\times10^3:7\times 10^4]$ & $0.077(2)$ \\
~~~~$0.1$ & $[5\times10^3:5\times 10^4]$ & $0.069(2)$ \\
~~~~$0.5$ & $[1\times10^2:1\times 10^4]$ & $0.028(2)$ \\
\hline
\end{tabular}
\end{table}
\par
The main panel of Fig.\ \ref{persistence_composite} presents the time dependence of the composite persistence probability $P_{\rm c}(t)$ for different values of $p_r$, obtained for the same system size $L=256$ considered throughout this work. The data clearly exhibit two distinct stages in the decay of $P_{\rm c}(t)$. At early times, a transient slow-decay regime is observed, the duration of which increases with decreasing $p_r$. In the long-time limit, the data display a power-law decay. However, unlike the total and spin-flip persistence probabilities discussed earlier, the slopes of the curves gradually decrease with increasing $p_r$, indicating a $p_r$-dependent persistence exponent $\theta_{\rm c}$. This observation suggests that, unlike the previous persistence measures, the asymptotic decay behavior of the composite persistence is not characterized by a universal exponent. Motivated by this behavior, we fit the long-time data for different values of $p_r$ using the power-law form given in Eq.\ \eqref{P_power_law}. The resulting estimates are summarized in Table\ \ref{tab1}.
\par
From the values presented in Table\ \ref{tab1}, it can be inferred that $\theta_{\rm c}$ approaches the value obtained for the total and spin-flip persistence as $p_r$ decreases. This is also clearly evident from the inset of Fig.\ \ref{persistence_composite} showing the plots (squares) of the same variation. At first sight, this appears to be a rather counter intuitive observation, since smaller values of $p_r$ imply that spin-exchange moves dominate over spin-flip moves. However, it is important to recall that the decay of the composite persistence requires the action of both microscopic processes on a given site. Consequently, for small values of $p_r$, the overall decay rate is governed by the bottleneck process, namely the comparatively rare spin-flip events. At early times and for small $p_r$, the evolution is primarily driven by spin-exchange moves. As a result, $P_{\rm c}(t)$ exhibits an initial regime with very little appreciable decay. However, after a certain period, the system reaches a stage where spin-exchange moves become progressively less effective because a majority of sites have already experienced such events. Thereafter, the subsequent decay of composite persistence is controlled predominantly by spin-flip events occurring on these sites. Consequently, the asymptotic behavior approaches that of persistence in a system evolving purely under nonconserved dynamics. As $p_r$ increases, conserved and nonconserved moves are accepted with comparable frequencies and become simultaneously active throughout the evolution. Hence, the decay rate of $P_{\rm c}(t)$ systematically deviates from that of conventional nonconserved persistence, resulting in a slower asymptotic decay.  We also include the dependence of the saturation value $P_{\rm c}^{\rm sat}$ on $p_r$. The observed trend is similar to that found for the total persistence in Fig.\ \ref{psat_vs_pr}(b).
\par
\begin{figure}[t!]
\centering
\includegraphics*[width=0.5\textwidth]{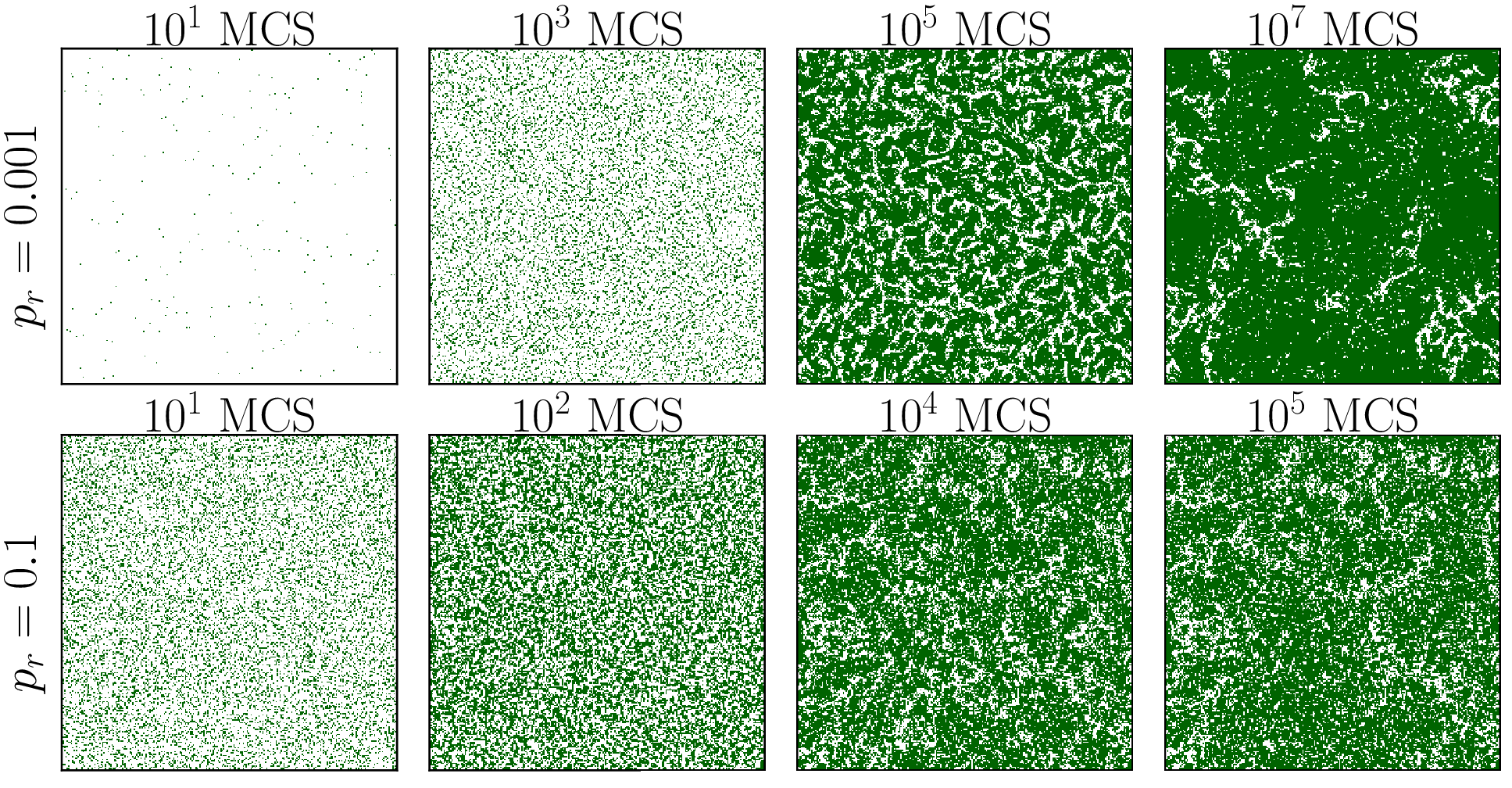}\\
\caption{\label{snap_composite} Time evolution snapshots of the composite persistence lattice for two different values of $p_r$. Only the non-persistent sites are marked in green. The snapshots correspond to the same system studied throughout.}
\end{figure}

Having developed an understanding of the behavior of the composite persistence probability, we next investigate the morphological properties of the corresponding persistence lattice. Representative time evolution snapshots of the composite persistence lattice for two different values of $p_r$ are presented in Fig.\ \ref{snap_composite}. Here we define the corresponding order parameter as $\rho_{\rm c}(\mathbf{x}_{\rm c},t)$, which takes the value $1$ if a site remains composite persistent at time $t$, and $0$ otherwise.
The snapshots at different times indicate the formation and growth of spatially correlated regions of non-persistent sites, which exhibit signatures of fractal morphology at late times. However, one also notices that for the larger value $p_r=0.1$, the growth process is considerably slower and the morphology does not evolve as far as that observed for smaller values of $p_r$. Consequently, the fractal structures observed at late times for the two values of $p_r$ appear markedly different. Examination of the corresponding time evolution snapshots for other values of $p_r$ (not shown here) also reveals a systematic variation of the fractal morphology with $p_r$.

\begin{figure}[b!]
\centering
\includegraphics*[width=0.5\textwidth]{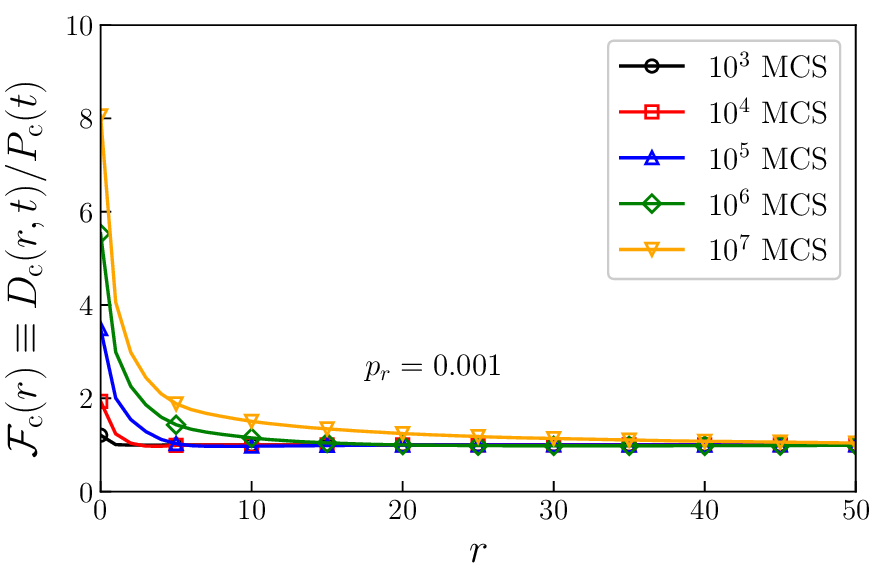}\\
\caption{\label{composite_corr} Plots of the scaling function $\mathcal{F}_{\rm c}(r)\equiv D_{\rm c}(r,t)/P_{\rm c}(t)$ at different times for $p_r=0.001$, demonstrating the development of an evolving characteristic length scale during the growth process. The system size is $L=256$.}
\end{figure}

\par
To further explore the morphological properties, we calculate the spatial correlation function of the composite persistence lattice at a given time as

\begin{equation}\label{corr_comp_pers_def}
 D_{\rm c}(r,t)=\frac{\langle \rho_{\rm c}(\mathbf{x}_{\rm c},t)\rho_{\rm c}(\mathbf{x}_{\rm c}+\mathbf{r},t)\rangle}{\langle \rho_{\rm c}(\mathbf{x}_{\rm c},t) \rangle}.
\end{equation}
Here again, by definition, $\langle \rho_{\rm c}(\mathbf{x}_{\rm c},t)\rangle=P_{\rm c}(t)$. Plots of $D_{\rm c}(r,t)/P_{\rm c}(t)$ at different times, presented in Fig.\ \ref{composite_corr}, clearly indicate the presence of a growing characteristic length scale $\ell_p^{\rm c}(t)$, similar to the behavior observed for the other persistence lattices considered earlier. This motivates us to construct an analogous scaling ansatz of the form
\begin{equation}\label{corr_comp_scaling_func}
 D_{\rm c}(r,t)=P_{\rm c}(t)\mathcal{F}_{\rm c}(y),
\end{equation}
where $\mathcal{F}_{\rm c}(y)$ is the corresponding scaling function and the scaling variable is defined as
\begin{equation}\label{scaling_y_comp}
 y=\frac{r}{\ell_p^{\rm c}(t)}.
\end{equation}
The decay exponent $\kappa=d-d_f^{\rm c}$ characterizing the small-$y$ behavior of $\mathcal{F}_{\rm c}(y)$ at late times can then be used to estimate the fractal dimension $d_f^{\rm c}$ through the asymptotic form 
\begin{equation}\label{comp_fy_behavior}
\mathcal{F}_{\rm c}(y)=\frac{D_{\rm c}(r,t)}{P_{\rm c}(t)}=
\begin{cases}
y^{-\kappa}, & y\ll 1,\\
1, & y\gg 1.
\end{cases}
\end{equation}
Subsequently, one can examine whether an analogous universal scaling relation, as embodied in Eq.\ \eqref{scaling_relation}, also holds for the composite persistence lattice.
\par
\begin{figure}
\includegraphics*[width=0.5\textwidth]{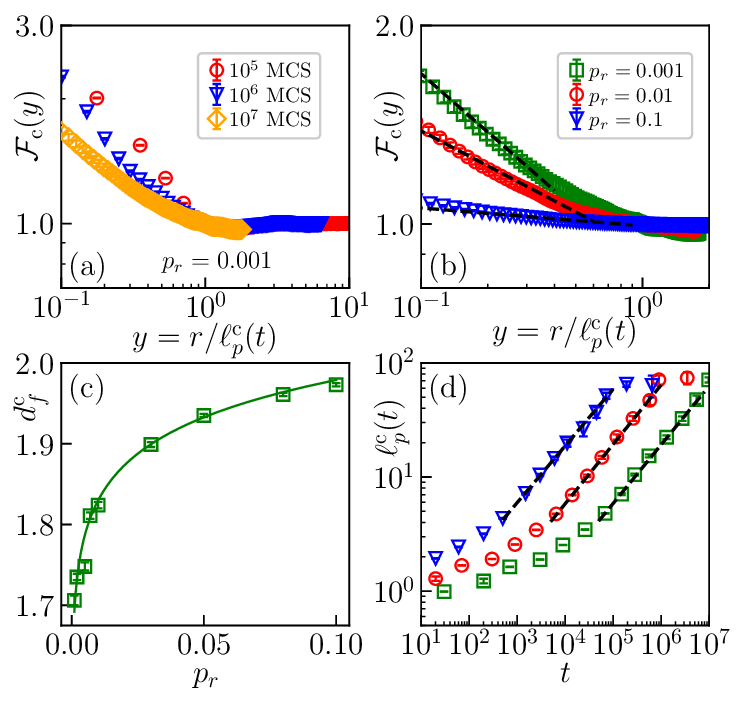}\\
\caption{\label{composite_length} (a) Double-logarithmic plots of the scaling function $\mathcal{F}_c(y)$ for $p_r=0.001$ at different times. (b) Corresponding double-logarithmic plots of $\mathcal{F}_c(y)$ for three different values of $p_r$, shown at their respective latest times. The dashed lines represent the corresponding fits using Eq.\ \eqref{comp_fy_behavior}. (c) Variation of the fractal dimension $d_f^{\rm c}$ of the composite persistence lattice as a function of $p_r$. The continuous curve represents a fit using the form given in Eq.\ \eqref{df_vs_pr}. 
(d) Double-logarithmic plots of the time dependence of the characteristic length scale $\ell_p^{\rm c}(t)$ for different $p_r$. The dashed lines represent the corresponding fits using Eq.\ \eqref{lp_comp_law}. In all the cases the considered system size is $L=256$.}
\end{figure}

Exemplary plots of $\mathcal{F}_c(y)$ for $p_r=0.001$ are presented in Fig.\ \ref{composite_length}(a). Similar to the case of total persistence, the data at different times do not exhibit a satisfactory collapse. Nevertheless, the small-$y$ behavior of the scaling function can still be used to estimate $\kappa$, and thereby the fractal dimension $d_f^{\rm c}$, by fitting the latest-time data of $\mathcal{F}_c(y)$ for different values of $p_r$ using the form given in Eq.\ \eqref{comp_fy_behavior}. The corresponding latest-time data for $\mathcal{F}_c(y)$ are shown in Fig.\ \ref{composite_length}(b) for three representative values of $p_r$. The plots clearly suggest that $\kappa$ decreases with increasing $p_r$. The fitting parameters obtained for a broader range of $p_r$ values are summarized in Table\ \ref{tab2}, further supporting this observation.

\par
The decreasing trend of $\kappa$ implies that $d_f^{\rm c}$ approaches the spatial dimension $d$ as $p_r \rightarrow 1$. This behavior is illustrated in Fig.\ \ref{composite_length}(c). At first sight, this observation appears rather surprising since, for conventional persistence in a system evolving purely under nonconserved dynamics ($p_r=1$), the persistence lattice is known to possess a significantly smaller fractal dimension. However, one should recall that for sufficiently small values of $p_r$, the late-time behavior of the composite persistence lattice is predominantly governed by the nonconserved spin-flip dynamics. Consequently, the resulting morphology inherits characteristics similar to those of the conventional persistence lattice, leading to a comparatively smaller value of $d_f^{\rm c}$. To quantify the variation of $d_f^{\rm c}$ with $p_r$, we perform a power-law fit using the form
\begin{equation}\label{df_vs_pr}
 d_f^{\rm c}=Mp_r^{\beta},
\end{equation}
which yields a rather small exponent, $\beta=0.034(5)$. The corresponding fit is shown by the continuous curve passing through the data points in Fig.\ \ref{composite_length}(c).

\begin{table}[b!]
\centering
\caption{Estimated values of $\kappa$ and the corresponding fractal dimension $d_f^{\rm c}$ of the composite persistence lattice, obtained by fitting the small-$y$ behavior of $\mathcal{F}_c(y)$ using Eq.\ \eqref{comp_fy_behavior} for different $p_r$. }
\label{tab2}
\begin{tabular}{llll} 
\hline
\hline
~~~$p_r$~~~~~~~~~&fitting $y$ range~~~~~~~~&~~~~~$\kappa$~~~~~~~~~~&$d_f^{\rm c}=d-\kappa$ \\
\hline
$0.001$ & $[0.05:0.40]$ & $0.294(6)$ & $1.706(6)$\\
$0.002$ & $[0.06:0.50]$ & $0.265(4)$ & $1.735(4)$\\
$0.005$ & $[0.06:0.60]$ & $0.252(5)$ & $1.748(5)$ \\
$0.007$ & $[0.06:0.65]$ & $0.189(4)$ & $1.811(4)$ \\
~~$0.01$ &$[0.07:0.70]$ & $0.176(4)$ & $1.824(4)$ \\ 
~~$0.03$ & $[0.07:0.70]$ & $0.101(3)$ & $1.899(3)$ \\
~~$0.05$ & $[0.07:0.75]$ & $0.065(2)$ & $1.935(2)$  \\
~~$0.08$ & $[0.08:0.85]$ & $0.039(2)$ & $1.961(2)$  \\
~~~~$0.1$ & $[0.09:0.90]$ & $0.028(2)$ & $1.972(2)$ \\
\hline
\end{tabular}
\end{table}
\par
Next, in Fig.\ \ref{composite_length}(d), we present the time dependence of the characteristic length scale $\ell_p^{\rm c}(t)$ for three different values of $p_r$. The overall behavior is similar to what is observed for the data of $\ell_p^{\rm t}(t)$ and $\ell_p^{\rm f}(t)$ presented in Fig.\ \ref{length_persistence}. To quantify the asymptotic power-law behavior, evident from the linear variation of the data on a double-logarithmic scale, analogous to Eq.\ \eqref{lp_law}, we employ the  ansatz 
\begin{equation}\label{lp_comp_law}
\ell_p^{\rm c}=A_{\rm c} t^{\alpha_{\rm c}}.
\end{equation}
Fitting the data shown in Fig.\ \ref{composite_length}(d) using the above form yields $\alpha_{\rm c}=0.515(3)$, $0.522(2)$, and $0.498(2)$ for $p_r=0.001$, $0.01$, and $0.1$, respectively. Similar fitting analyses performed for other values in the range $p_r \in [0.001,0.1]$ produce estimates within $\alpha_{\rm c}\in[0.485,0.535]$. These results strongly suggest an asymptotic power-law growth consistent with the LCA prediction.
\par
Finally, having obtained estimates of the exponents $\theta_{\rm c}$, $d_f^{\rm c}$, and $\alpha_{\rm c}$, we note that among these quantities only the growth exponent $\alpha_{\rm c}$ exhibits a universal behavior, essentially independent of $p_r$. In contrast, both $\theta_{\rm c}$ and $d_f^{\rm c}$ vary monotonically with $p_r$. Importantly, the values reported in Tables\ \ref{tab1} and \ref{tab2} indicate that, unlike the total and spin-flip persistence, the composite persistence does not satisfy a scaling relation analogous to Eq.\ \eqref{scaling_relation}. In other words,
\begin{equation}
 d-d_f^{\rm c} \ne \frac{\theta_{\rm c}}{\alpha_{\rm c}}.
\end{equation}

\section{Conclusion}\label{conclusion}
We have investigated the persistence properties of the Ising model in spatial dimension $d=2$ during phase ordering under competing conserved spin-exchange and nonconserved spin-flip dynamics by means of  Monte Carlo simulations at zero temperature. Along with the conventional definition of persistence (which we refer to as the total persistence), we have also studied two additional persistence measures, viz., the spin-flip persistence and the composite persistence in order to isolate the role of individual microscopic processes and characterize dynamically inactive regions unaffected by any of the microscopic moves. At the outset, the presence of dynamical scaling in the original spin lattice was verified using the two-point equal-time spin-spin correlation function. The corresponding characteristic length scale was found to obey the LCA law with a growth exponent $\alpha \approx 1/2$.

\par
Both the total and spin-flip persistence probabilities were found to exhibit asymptotic power-law decay with an exponent $\theta_i \approx 0.225$, independent of the relative occurrence probability $p_r$ of the spin-flip moves. This value was found to be consistent with the persistence exponent corresponding to purely nonconserved dynamics. In both cases, a transient regime preceding the asymptotic behavior was observed. During this regime, the total persistence exhibited a featureless decay, whereas the spin-flip persistence remained nearly constant. The duration of this transient regime was quantified in terms of the crossover time $\tau_c$, defined as the time at which the spin-flip persistence probability $P_{\rm f}(t)$ begins to decay appreciably. Numerical estimates suggested a power-law scaling $\tau_c \sim p_r^{-1.07}$. Combining this crossover behavior with the asymptotic decay,  $P_{\rm f}(t)$ was found to obey the dynamical scaling relation
$
P_{\rm f}(t)=P_{\rm f}^{\rm sat}f\left(\frac{t}{\tau_c}\right),
$
with $f(y)\sim y^{-\theta_{\rm f}}$ in the asymptotic limit, irrespective of $p_r$.
\par
We have also explored the morphological properties of the corresponding persistence lattices. Spatial correlation analyses revealed the development of fractal morphologies characterized by a growing length scale. The corresponding growth exponent $\alpha_i\approx 1/2$ was found to remain consistent with the LCA prediction, while the estimated fractal dimensions $d_f^i\in [1.51,1.6]$ remained close to values reported for systems evolving purely under nonconserved dynamics. Furthermore, the scaling relation $d-d_f^i=\theta_i/\alpha_i$ was found to remain valid for both persistence measures despite the simultaneous presence of conserved and nonconserved microscopic dynamics.

\par
In contrast, the composite persistence probability was found to exhibit qualitatively different behavior as a function of $p_r$. Even though the asymptotic power-law decay was observed, the corresponding persistence exponent $\theta_{\rm c}$ displayed a systematic dependence on $p_r$. The evolving morphology of the composite persistence lattice was also found to depend strongly on $p_r$. Consequently, the corresponding fractal dimension $d_f^{\rm c}$ varied monotonically with $p_r$. Despite the growth exponent $\alpha_{\rm c}$ remaining close to the universal value $\alpha=1/2$, no analogous scaling relation connecting $\theta_{\rm c}$, $d_f^{\rm c}$, and $\alpha_{\rm c}$ was found to hold.

\par
The results presented here have demonstrated that conventional persistence properties retain universal features even under mixed dynamics, whereas the composite persistence captures nontrivial consequences arising from the competition between microscopic conserved and nonconserved dynamic moves. The observed breakdown of the scaling relation for composite persistence suggests the emergence of additional mechanisms governing the evolution of dynamically inactive regions.
\acknowledgments
S.T. and S.M. acknowledge funding by the Anusandhan National Research
Foundation (ANRF), Government of India, through a Ramanujan Fellowship (File No.\ RJF/2021/000044). The authors also acknowledge the use of ChatGPT (OpenAI) for assistance in polishing the English language to improve the readability of the manuscript.

%
\end{document}